\documentclass[
reprint,
superscriptaddress,
bibnotes,
amsmath,amssymb,
aps,
prl,
]{revtex4-1}

\usepackage{graphicx}% Include figure files
\usepackage{dcolumn}% Align table columns on decimal point
\usepackage{bm}% bold math
\usepackage{color}
\usepackage{umoline}

\begin{document}

\preprint{APS/123-QED}

\title{Eigenmodes of a quartz tuning fork and their application to photo-induced force microscopy}
% Force line breaks with \\

\author{Bongsu Kim}
\affiliation{Department of Chemistry, University of California, Irvine, California 92697, USA}
\author{Junghoon Jahng}
\affiliation{Center for Nanometrology, Korea Research Institute of Standards and Science, Daejeon 304-340, South Korea}
\author{ Ryan Muhammad Khan}
\affiliation{Department of Chemistry, University of California, Irvine, California 92697, USA}
\author{ Sung Park}
\affiliation{Molecular Vista, 6840 Via Del Oro, San Jose, California 95119, USA}
\author{Eric O. Potma}
\email{Correspondence: epotma.uci.edu}
\affiliation{Department of Chemistry, University of California, Irvine, California 92697, USA}

\date{\today}% It is always \today, today,

\begin{abstract}

We examine the mechanical eigenmodes of a quartz tuning fork (QTF) for the purpose of facilitating its use as a probe for multi-frequency atomic force microscopy (AFM). We perform simulations based on the three-dimensional finite element method (FEM) and compare the observed motions of the beams with experimentally measured resonance frequencies of two QTF systems. The comparison enabled us to assign the first seven asymmetric eigenmodes of the QTF. We also find that a modified version of single beam theory can be used to guide the assignment of mechanical eigenmodes of QTFs. The usefulness of the QTF for multi-frequency AFM measurements is demonstrated through photo-induced force microscopy (PiFM) measurements. By using the QTF in different configurations, we show that the vectorial components of the photo-induced force can be independently assessed, and that lateral forces can be probed in true non-contact mode.

%\begin{description}
%\end{description}
\end{abstract}

\maketitle

%\tableofcontents

\section{\label{sec:level1}I. Introduction}

The quartz tuning fork (QTF) is a popular probe used in scanning probe microscopy (SPM). The QTF can be used as a piezoelectric resonator, which features a high quality factor, excellent stability and a small oscillation amplitude. These attributes have made it possible to collect topographic atomic force microscopy (AFM) images with subatomic 
resolution in ultrahigh vacuum~\cite{sci348}. Compared to a cantilevered tip, the QTF is less prone to the unwanted the jump-to-contact problem under ambient 
conditions~\cite{rsi85}. In addition, the QTF provides a way to investigate the sample through oscillatory motions in the lateral plane parallel to the sample, allowing precise shear force measurements that are more difficult to perform with cantilevers~\cite{nc6}. 

The QTF consists of two quartz beams with applied electrodes. The motions of the QTF can be driven both electrically and mechanically. Electrical driving is accomplished through the electrodes for exciting a selected eigenmode of the system, commonly an asymmetric in-plain bending mode~\cite{sen148}. Mechanical driving can couple effectively to other modes as well, including out-of-plane and torsional modes, which are not easily accessible through electrical excitation. Some of the representative eigenmodes of a QTF system are shown in Figures \ref{f1} and \ref{f2}. It is clear that a QTF exhibits a series of useful eigenmodes that can be utilized in multi-frequency AFM, which is based on the notion that multiple modes of the probe are exploited for examining different tip-sample interactions simultaneously~\cite{NatTech12}.  The concept of multi-frequency AFM has been successfully implemented with cantilever probes, for instance in Kelvin probe force microscopy (KPFM)~\cite{apl58}, mechanical stiffness and damping 
measurements~\cite{prl111}, and photo-induced force microscopy (PiFM)~\cite{prb90}. Because of its rich set of distinct eigenmodes, the QTF is ideally suited as a probe for multi-frequency AFM. Yet, multi-frequency implementations of tuning forks in scan probe microscopy are scarce.

Most experiments use the first fundamental asymmetric in-plane bending motion of the QTF in AFM applications. The QTF has been used in both tapping and in shear mode, but this is typically accomplished by mounting the tuning fork in distinct orientations while still using the fundamental in-plane bending mode for readout~\cite{prl111}. One of the complications of using multiple modes in the QTF simultaneously is that assigning the observed resonances of the system to specific eigenmotions is not straightforward. Whereas single beam theory has been successfully used to identify the nature of the modes in cantilever beams~\cite{prb79}, such approaches are more involved in the case of the QTF, which consists of two coupled beams. 

To benefit from the versatility of QTFs in multi-frequency AFM, assigning the accessible eigenmodes of the probe is a necessary step. More insight in the QTF motions can be obtained from three-dimensional finite element methods (FEM). In this paper, we examine the first seven asymmetric eigenfrequencies of two QTF systems through FEM simulations and compare these findings with the experimentally measured resonance frequencies of the probe. Based on the thus acquired simulation and measurements, we show that a modified single beam theory for asymmetric uncoupled QTF motion can be used to achieve good approximations for the observed resonance frequencies. In addition, having fully characterized the accessible eigenmodes of the probe, we apply the QTF to multi-frequency PiFM by using the first in-plane mode for registering the photo-induced force and the second in-plane mode for active feedback. Through these measurements, we demonstrate that the quartz tuning fork can be used to register not only the forces induced by a tightly focused laser beam in the axial (tapping) direction, but also in the lateral (shear force) direction. This latter capability is unique to the QTF and is not easily attained with a cantilevered tip.

%\\\\\\\\\\\\\\\\\\\\\\\\\\\\\\\\\\\\\\\\\\\\\\\\\\\\\\\\\\\\\\\\\\\\\\\\\\

\section{\label{sec:level1}II. Eigenmodes of the QTF }
\subsection{\label{sec:level1}1. Simulated resonances}

\begin{figure}[bth]\centering
\includegraphics[scale=0.43]{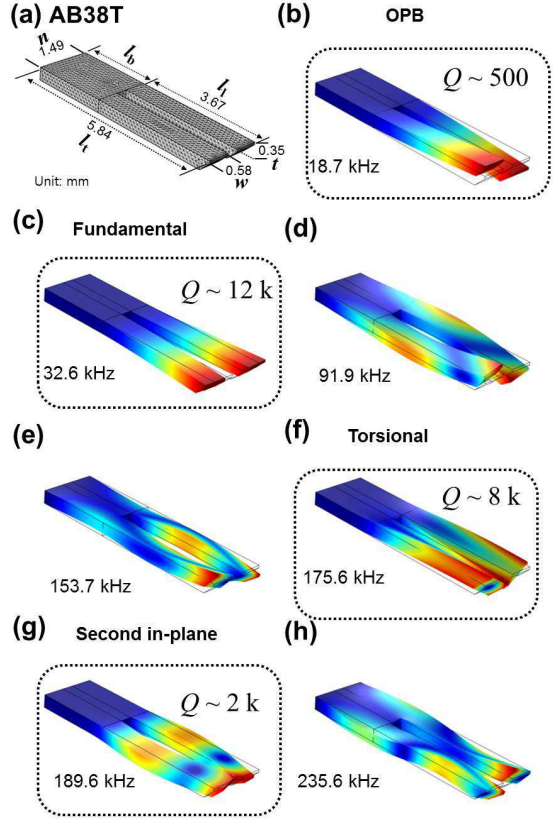}
\caption{\label{f1} FEM simulations of the QTF asymmetric modes of the AB38T tuning fork.
(a) Dimensions of the tuning fork (unit: mm).
(b) Out-of-plane bending (OPB) mode.
(c) In-plane bending (fundamental) mode.
(d, e) Coupling mode.
(f) Torsional mode.
(g) Second in-plane bending mode.
(h) Coupling mode. The quality factors ($Q$) are measured value.  
} 
\end{figure}

\begin{figure}[bth]\centering
\includegraphics[scale=0.43]{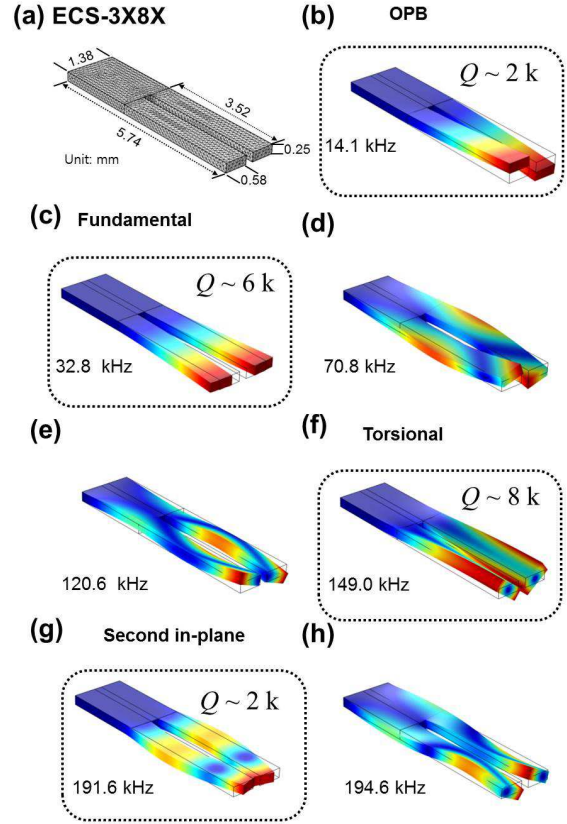}
\caption{\label{f2} FEM simulations of the QTF asymmetric modes of the ECS-3X8X tuning fork.
(a) Dimensions of the tuning fork (unit: mm).
(b) Out-of-plane bending (OPB) mode.
(c) In-plane bending (fundamental) mode.
(d, e) Coupling mode.
(f) Torsional mode.
(g) Second in-plane bending mode.
(h) Coupling mode. The quality factors ($Q$) are measured value.  
} 
\end{figure}

\subsection{\label{sec:level1}1. Simulation of eigenfrequencies}
The QTF motions and eigenfrequencies were simulated with three-dimensional FEM using COMSOL MultiPhysics 5.0 software package and the Solid Mechanics module. The geometry is depicted in Figures \ref{f1}(a) and \ref{f2}(a), for the AB38T and ECS-3X8X tuning forks, respectively.
The structure is composed of three subdomains; one body (length$\times$width$\times$thickness, $l_{\rm b} \times n \times t$) and two legs ($l_{\rm l} \times w \times t$). The substrate material is quartz with a density of 2649 kg/m$^3$, a Young's modulus of 76.5 GPa and Poisson's ratio set to 0.228~\cite{hb}.
The model is meshed in tetrahedral blocks with a fine element size. Eleven eigenfrequencies were found for the AB38T-QTF, and twelve eigenfrequencies were found for the ECS-3X8X structure under 200 kHz. The first seven asymmetric modes are shown in Figures \ref{f1} and \ref{f2}. Specifically, in panels (b) of both Figures the out-of-plane bending (OPB) are shown. In panels (c), the fundamental in-plane bending modes are depicted, whereas panels (f) and (g) show the torsional and the 2nd in-plane bending modes, respectively. In addition, panels (d), (e), and (h) represent coupled beam modes, which are not observed for a single beam.

\subsection{\label{sec:level1}2. Experimental resonances}

\begin{figure}[bth]\centering
\includegraphics[scale=0.6]{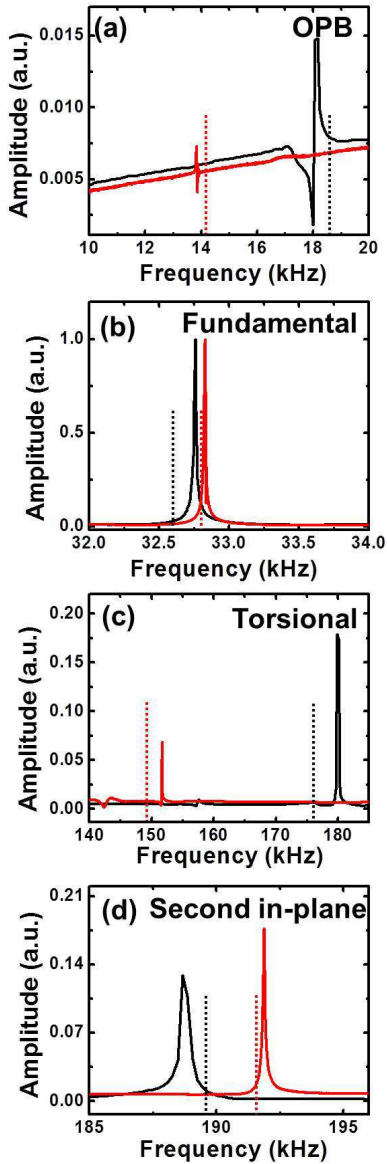}
\caption{\label{f3} The amplitude displaced as a function of the QTF driving frequency. Black (red) solid line presents the measured amplitude of AB38T (ECS-3X8X),
whlie the black (red) line is the eigenfrequency of the AB38T QTF computed by COMSOL Multiphysics 5.0.
(a) Out-of-plane bending (OPB) mode
(b) Fundamental (first in-plane bending) mode
(c) Torsional mode
(d) Overtone (second in-plane bending) mode
}
\end{figure}

We used a commercial AFM system (VistaScope, equipped with tuning fork head) to measure the eigenfrequencies of the QTF system experimentally. In this experiment, we examined the bare QTF without a tip mounted to it. The experiments were performed under ambient conditions. The QTF was mechanically driven by a piezo-electric element brought into contact with the tuning fork, and the examined driving frequency was swept from 10 to 200 kHz. The motions of the QTF were measured electronically, by detecting the piezo-electric response of the quartz due to the frequency-dependent deformation of the material. Figure \ref{f3} presents the results of the measurement for both QTFs. The experiments are compared with the resonance frequencies of the simulation, as indicated by the dotted lines. As can be gleaned from the Figure, the predicted eigenfrequencies are close to the measured resonance frequencies to within a few kHz. The slight mismatch between experiment and simulation can be largely explained by the relative uncertainty of $\pm10\;\mu$m in the exact dimensions of the QTZ beams.

\subsection{\label{sec:level1}3. Single beam approximation}

\begin{figure}[bth]\centering
\includegraphics[scale=0.43]{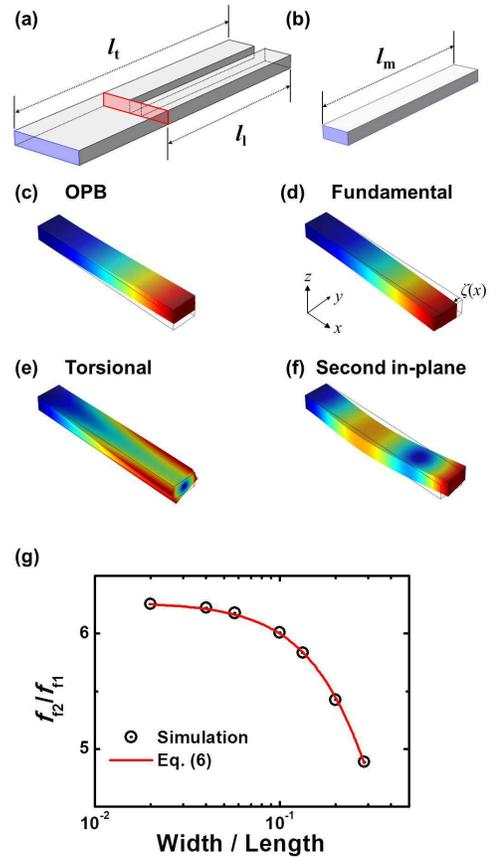}
\caption{\label{f4} Single beam approximation.
(a) Definition of the total length $l_{\rm t}$ and leg length $l_{\rm l}$. The blue plane presents the fixed constraint of the QTF.
(b) The effective length $l_{\rm m}$ in the single beam approximation. The blue plane presents the fixed constraint of the single beam.
(c)--(f) Single beam motion calculated by COMSOL Multiphysics 5.0.
(c) Out-of-plane bending(OPB) mode
(d) Fundamental (first in-plane bending) mode. $\zeta(x)$ is the displacement from a neutral point of beam.
(e) Torsional mode
(f) Second in-plane bending mode
(g) The ratio of second in-plane bending eigenfrequency to fundamental eigenfrequency depending on thickness/length of a single beam.
The black data was obtained by COMSOL Multiphysics 5.0, and it was fitted by Eq. (\ref{eq_06}).
}
\end{figure}

The FEM simulations confirm that QTFs exhibit various mechanical eigenmodes. Except for the fundamental resonance, however, it is not always straightforward to compare the experimental resonance frequencies with the ones found in the simulation, because some eigenmotions appear close to one another in the same frequency range. This issue is more pertinent to the case of the QTF than for the case of cantilevers, because of the presence of coupled beam motions in QTFs, which introduce more resonances. Moreover, the sequence in which the various eigenmodes appear as a function of driving frequency may change for small variations in the beam dimensions. Consequently, slight uncertainties in the QTFs dimensions can lead to the wrong assignment.

In this regard, it would be desirable to have additional clues that can help guide the assignment. In this Section we show that it is possible to use a modified version of single beam theory to gain insight in the resonance frequencies of the asymmetric and uncoupled eigenmodes of the QTF. First, we make the approximation that coupling between the beams does  not significantly affect the resonance frequency of the eigenmode. This is a rather stringent requirement, but measurements have shown that the shift due to coupling is under most conditions minimal~\cite{bj5, apl91}. Within this assumption, the resonance frequencies of a single beam (dimension $l \times w \times t$ ) can be easily calculated as follows\cite{landau, butt, jap92},

\begin{equation}
\label{eq_01}
f_{\rm o} =\frac{1.01489}{2\pi}\frac{t}{l^2}\sqrt{\frac{E}{\rho}},
\end{equation}

\begin{equation}
\label{eq_02}
f_{\rm f1} =\frac{1.01489}{2\pi}\frac{w}{l^2}\sqrt{\frac{E}{\rho}},
\end{equation}

\begin{equation}
\label{eq_03}
f_{\rm t} =\frac{1}{4l}\sqrt{\frac{GK}{\rho I}},
\end{equation}
where $E$ is Young's modulus, $\rho$ is the material density, $G$ is a shear modulus, $K$ is a geometric function of the cross section, $I$ is the polar moment of inertia about the axis of rotation, and $f_{\rm o}$,  $f_{\rm f1}$, $f_{\rm t}$ are the OPB, first fundamental and torsional eigenfrequencies, respectively. 
For a rectangular beam, $K$ and $I$ are expressed as\cite{landau, jap99},

\begin{equation}
\label{eq_04}
K\approx\frac{1}{3}wt^3\left(1.0-0.63\frac{t}{w}+0.052\frac{t^3}{w^3},
\right)
\end{equation}

\begin{equation}
\label{eq_05}
I=\frac{1}{12}(tw^3+wt^3).
\end{equation}
Since the Young's modulus, shear modulus and density are material characteristics, once the beam dimensions ($l$, $w$, $t$) are determined with accuracy, the three resonance frequencies are readily calculated. However, with reference to Figure \ref{f4}(a) and (b), there is ambiguity in determining the effective length of the beam. In the QTF, the two beams are connected to the body, indicated in panel (a) by the red rectangle. It is incorrect to assume that at the position of the red rectangle the beam is completely motionless, as would be the case for the single beam, which is fixed at the position of the blue rectangle as shown in panel (b). Since the motion of each beam extends beyond length $l_{\rm l}$, we may assume that the effective length of the beams is in between $l_{\rm l}$ and $l_{\rm t}$. To overcome the uncertainty in the effective length of the beam, we use the known information about the thickness and width of the beam in combination with information about the fundamental in-plane bending mode. Since the $f_{f1}$ mode can be assigned with certainty, we may use the measured value of the resonance frequency of the mode in Eq. (\ref{eq_02}) to determine the modified length ($l_{\rm m}$) of the single beam. Using the modified length, the resonance frequencies of the OPB and torsional eigenmodes can then be predicted with Eqs. (\ref{eq_01}) and (\ref{eq_03}). For comparison, the simulated asymmetric and uncoupled motions associated with the OPB and torsional modes of a single beam are shown in Figure \ref{f4}(c) and \ref{f4}(e), respectively. 

The second mechanical resonance $f_{f2}$ of the single beam is depicted in Figure \ref{f4}(f). Unlike for the OPB and torsional modes, it is more involved to obtain a quick estimate for this resonance. In case of the AFM-cantilever, the second mechanical resonance can be obtained by multiplying the $f_{f1}$ frequency by 6.27~\cite{butt, nnt7}. This procedure is valid if the beam length is much larger than its width, i.e. $l_{\rm l}>>w$. The latter is a good approximation for a cantilever beam, but not necessarily for a QTF, because the ratio $w/l_{\rm l}$ is substantially larger for the QTF beam. To retrieve a relation between the $f_{f1}$ and $f_{f2}$, we have performed simulations of the ratio of $f_{\rm f2}$ over $f_{\rm f1}$ for various dimensions of the QTF beam. In these simulations, the beam length was varied between $3.5w$ to $50w$ for a given beam width. The simulations yielded identical results for beam widths in the range of 10--20 $\mu$m, which are typical values for commercially available QTFs. The results are given by the open circles in the plot of Figure \ref{f4}(g). The graph clearly shows that the $f_{f2}$ is about 6.27 times higher than $f_{\rm f1}$ if $f_{\rm f2}/f_{\rm f1}<10^{-2}$, and that the ratio decreases for larger ratios (relatively thicker beams). 

The simulated trend shown in Figure \ref{f4}(g) is invariable to the thickness and beam width for practical beam dimensions. Therefore, the observed trend can be used for predicting the frequency of the second mechanical resonance if $f_{\rm f1}$ is known. The relation between $f_{\rm f1}$ over $f_{\rm f2}$ can also be obtained by fitting the simulated results with the following expression:
\begin{equation}
\label{eq_06}
\frac{f_{\rm f2}}{f_{\rm f1}}=6.27\left[1+\left(\frac{w}{l}\right)^a\right]^b,
\end{equation}
The fit shown in Figure \ref{f4}(g) was obtained with fitting parameters $a=1.695$ and $b=-2.224$. The formalism presented in this section can thus be used to obtain reasonable estimates for the eigenfrequencies of the fundamental,OPB, torsional and second mechanical resonance without resorting to full FEM simulations. Table \ref{table:1} presents the measured resonance frequencies, the FEM simulated frequencies and the estimates based on the formalism laid out through Eqs. (\ref{eq_01})--(\ref{eq_06}).

\begin{table}[h!]
\centering
 \begin{tabular}{c c c c c} 
 \hline
  & OPB & Fundamental & Torsional  & 2$^{nd}$ in-plane\\ [0.5ex] 
 \hline
Measur.&18.0(13.9)&32.8(32.8)&180.0(151.7)&188.7(191.9)\\ 
 QTF sim. &18.7(14.1) &32.6(32.8)&175.6(149.0)&189.6(191.6)\\
 Beam sim. &19.7(14.1)&32.3(32.3)&180.3(150.6)&185.6(185.6)\\
 Calculation &19.8(14.2)&32.8(32.8)&179.5(148.3)&188.8(188.8) \\ [1ex] 
 \hline
 \end{tabular}
\caption{Comparison of eigenfrequencies for the AB38T(ECS-3X8X) quartz tuning fork. Units are in kHz.}
\label{table:1}
\end{table}

%\\\\\\\\\\\\\\\\\\\\\\\\\\\\\\\\\\\\\\\\\\\\\\\\\\\\\\\\\\\\\\\\\\\\\\\\\\

\section{\label{sec:level1}III. Application: multi-frequency PiFM using a QTF}

\subsection{\label{sec:level1}1. Photo-induced Force}
We have utilized the QTF eigenmodes for probing the force experienced by the tip due to the presence of a tightly focused laser beam at a glass/air interface, see Figure \ref{f5}(a). The PiFM technique is based on detecting electromagnetically induced forces between the tip and the sample. A detailed explanation of the forces at play in PiFM within the dipole approximation is given in ref~\cite{prb90}. Here we briefly review the essentials of the technique for the purpose of demonstrating the usefulness of the QTF for detecting tip-sample interactions through selected mechanical eigenmodes of the tuning fork. Within the dipole approximation, the time-averaged Lorentz force experienced by the tip can be written as~\cite{prb90, novotny}:
\begin{equation}
\label{eq_07}
\langle{\bf{F}}\rangle=\sum_{s}
\left( 
\left<\frac{\alpha'_{ss}}{2}{\rm Re}\{E_{s}^{*}\nabla E_{s}\}\right>
+\left<\frac{\alpha''_{ss}}{2}{\rm Im}\{E_{s}^{*}\nabla E_{s}\}\right>
\right),
\end{equation}
here $\alpha'_{ss}$ and $\alpha_{ss}''$ are real and imaginary parts of the tip's polarizability ($\alpha_{ss}=\alpha'_{ss}+i\alpha_{ss}''$). The polarizability is a tensor, but for simplicity we shall here ignore the off-diagonal terms, i.e. the induced dipole direction is along the direction of the incident electric field ($s=x, y, z$). $E_s$ is a monochromatic electric field component in the near field polarized in the $s$-direction. When the tip is brought close to an interface, the electromagnetic field establishes a coupling between the induced dipole at the tip and its image dipole in the glass substrate. In our experimental scenario, the incident electric field $E_{0s}$ is an $x$-polarized field that is tightly focused by a high numerical aperture lens. Because of the tight focusing conditions, we expect a dominant $x$-polarized field $E_{0x}$, but also substantial portions of $y$ and $z$-polarized field components, $E_{0y}$ and $E_{0z}$, respectively~\cite{novotny,wolf}. To simplify the description we will assume that the fields near the focal plane exhibit minimum phase variations on the nm scale, in which case the field components can be approximated as real~\cite{novotny,Jahng2014}. Under these conditions, the time-averaged photo-induced force component can be summarized as:

\begin{equation}
\label{eq_08}
\langle{\bf{F}}\rangle_{s}
=
\frac{1}{4}\frac{\partial}{\partial s}\;
\bold{p_r}\cdot\bold{E}
+
\frac{\alpha''_{xx}}{2}E_{0x}^2k_z\delta_{zs}
\end{equation}
where $\bold{p_r}$ is the real part of the induced tip's dipole with the components $p_{s}=\alpha'_{ss}E_{0s}$, and $\delta_{ss'}$  is the Kronecker delta function.
Along the $x$-direction, the force manifests itself as a gradient force. In the $z$-direction, the force includes a gradient force, represented by the first term on the right hand side of Eq. \ref{eq_08}, and a scattering force, denoted by the second term. In the experiments that follow, we have implemented a side-band coupling scheme, which suppresses the contributions from the scattering force \cite{Jahng2016}. In this configuration, the measured signal is proportional to the gradient of the photo-induced force\cite{Jahng2016}. 
Therefore, the driven amplitude $A_{\rm{s}}$ of the probe by the photo-induced force along the $s$-direction is proportional to the second derivative of $\bold{p_r}\cdot\bold{E}$ as:

\begin{equation}
\label{eq_09}
A_{\rm{s}}
\propto
\left|
\frac{\partial^{2}}{\partial s^{2}}\;
\bold{p_r}\cdot\bold{E}
\right|
\end{equation}

%Because $|E_{0x}|^2$ is significantly larger than $|E_{0y}|^2$ and $|E_{0z}|^2$, we may expect that the lateral force (shear force) is dictated by the $x$-polarized incident field component. Along the axial direction (tapping force), we expect the $|E_{0z}|^2$ to dominate the photo-induced force, because $E_{0z}>E_{0y}$.%

\subsection{\label{sec:level1}2. Methods}

\begin{figure}[bth]\centering
\includegraphics[scale=0.43]{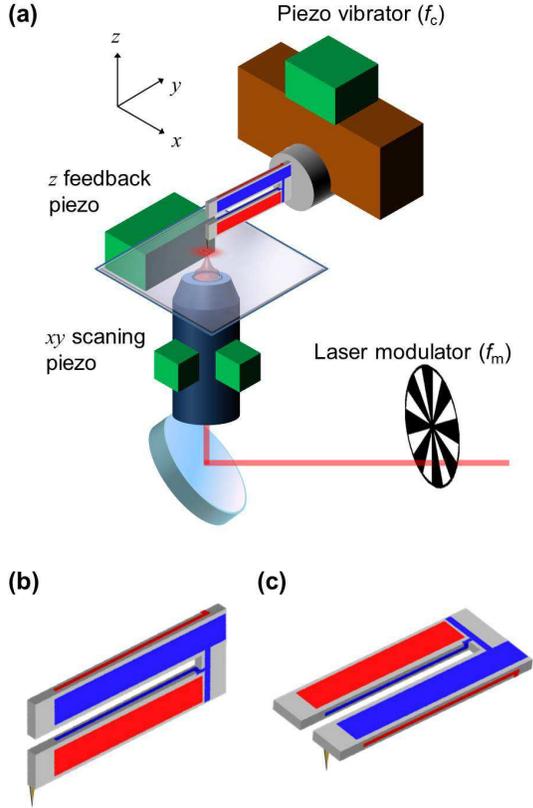}
\caption{\label{f5} (a) Diagram of experimental setup. (b) Orientation of the QTF and tip when measuring in tapping mode, using the in-plane bending mode. (c) Orientation of the QTF and tip when probing the sample in shear force mode, using the in-plane bending mode.
 }
\end{figure}

The experimental setup is shown in Figure \ref{f5}(a). In the experiments discussed here, the AB38T QTF was used. We have examined the PiFM signals for two different orientations of the tuning fork, as explained in Figures \ref{f5}(b) and \ref{f5}(c). For the configuration shown in \ref{f5}(b), the fundamental in-plane bending mode of the QTF probes the photo-induced tapping force, whereas the same mode in \ref{f5}(c) probes the photo-induced shear force. 

We have used the second in-plane mode for AFM feedback, whereas the first fundamental mode was used to probe the photo-induced force. The experiments were conducted with a femtosecond ti:sapphire laser as the pulsed light source (MaiTai, Spectra-Physics), which delivered 200 fs at a center wavelength of 809 nm with a pulse repetition rate of 80 MHz. The sideband-coupling scheme was used for detecting the PiFM signal. For this purpose, the laser beam was amplitude modulated with an acoustic optic modulator at frequency $f_m$, which was set as the sum of the fundamental ($f_{01}$) and the second in-plane eigenfrequency ($f_{02}$)~\cite{Jahng2016}. The sideband-coupled PIFM signal was obtained by demodulating the QTF response at $f_m-f_{02}$, which coincided with the fundamental resonance. The laser beam was focused by an NA $=$ 1.40 oil immersion objective. The incident laser power before the objective lens was on the order of 0.1 mW. A gold-coated silicon tip from a cantilever was attached (SICONGG, Applied NanoStructures) to the QTF. The sample consisted of a borosilicate glass slide (0.17 mm thickness). In the PiFM experiments, the objective was scanned in the $xy$ plane while the tip was held in place laterally, thus producing images of the focal field as detected by the photo-induced force. For the tapping mode measurements, the average tip-substrate distance was set to 2 nm with an oscillation amplitude of 1.5 nm. In the shear force measurements, the average tip-substrate distance was similar with a slightly larger oscillation amplitude of 2 nm.

\subsection{\label{sec:level1}3. Results and Discussion}

\begin{figure}[bth]\centering
\includegraphics[scale=0.43]{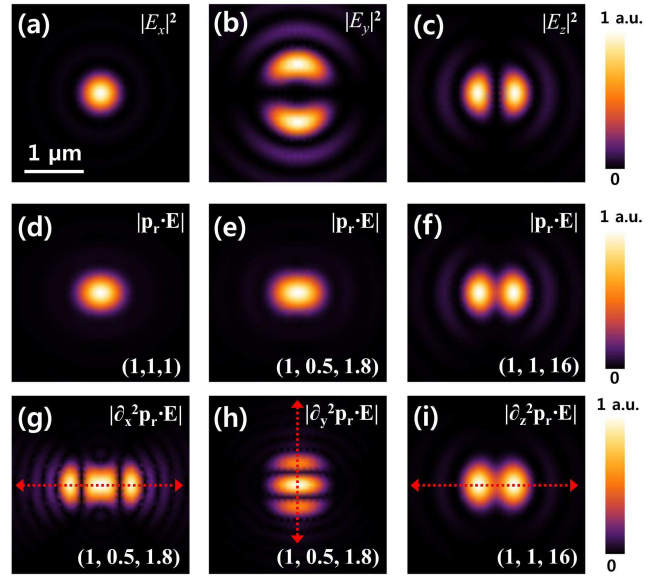}
\caption{\label{f6} Simulated PiFM maps of a tightly focused laser beam near a glass/air interface. 
 (a)--(c) Spatial intensity distributions of $E_x$(a), $E_y$(b) and $E_z$(c) in the focal plane.
(d)--(f) Magnitude of $\mid\bold{p_r}\cdot\bold{E}\mid$ for the different weights of the polarizability components. The numbers in the round bracket denote the (real) polarizability ratios as follows: ($\alpha'_{xx}/\alpha'_{xx}$, $\alpha'_{yy}/\alpha'_{xx}$, $\alpha'_{zz}/\alpha'_{xx}$). (g)--(i) The second derivative of $\mid\bold{p_r}\cdot\bold{E}\mid$, obtained from the panels (d)--(f), respectively. The profiles along the red dotted arrows are compared with the measured value in Fig. 7. Image size is 3 $\mu$m $\times$ 3 $\mu$m.}
\end{figure}

The computed intensity of a tightly focused laser beam is shown in Figures \ref{f6}(a)--(c) for each of the polarization components.
In Figure \ref{f6}(a), the $|E_{0x}|^2$ distribution is plotted, showing the Gaussian-type profile of the incident field.  The $y$ and $z$-components of the intensity distribution are given in panel \ref{f5}(b) and (c), revealing the characteristic double-lobed pattern of the $y$ and $z$-polarized focal field contributions. The force measurements are expected to follow the expression in Eq. \ref{eq_09}, which depends on the sum of the different the components of $\mathbf{p_r\cdot E}$. Figures \ref{f6}(d)--(f) show the spatial distribution of $\mathbf{p_r\cdot E}$, plotted for various selected weights of the components of the (real) polarizability. For an isotropic polarizability, i.e. $\alpha_{xx}=\alpha_{yy}=\alpha_{zz}$, the resulting spatial distribution resembles a Gaussian-like profile. In case the tip exhibits a higher polarizability along the $z$-direction, the system is more sensitive to the $E_{0z}$ component of the field and a characteristic double-lobed pattern is expected~\cite{Tamma2015}, as shown in panel \ref{f6}(f).

The second derivative of the patterns along $x$ and $y$ are shown in panels \ref{f6}(g) and \ref{f6}(h), respectively. These are representative patterns for the case when the field distribution is probed through photo-induced forces in the lateral direction, as measured in the shear force mode of Fig. \ref{f5}(c). Figure \ref{f6}(i) shows the second derivative of the pattern of panel \ref{f6}(f), representative for the case when the field distribution is probed in tapping mode, which is dominated by the $z$-polarized part of the focal field. This comparison makes it clear that very different PiFM maps can be expected in the tapping and shear force modes: the tapping mode is more sensitive to directional derivatives of the $E_{0z}$ component, producing a double lobed PiFM pattern, whereas the shear force mode is sensitive to the directional derivative of the total field along the lateral coordinate, giving rise to a characteristic triple-lobed PiFM image.

\begin{figure}[bth]\centering
\includegraphics[scale=0.43]{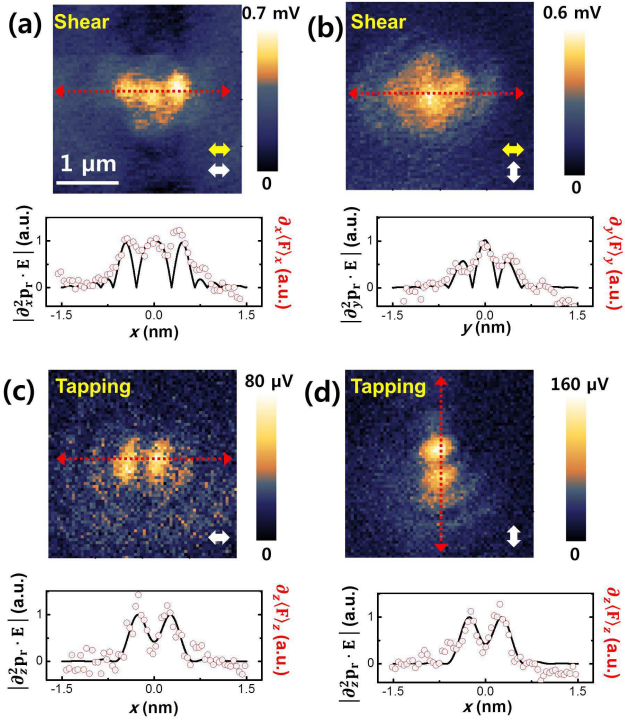}
\caption{\label{f7} Measured PiFM amplitude maps of a tightly focused laser beam  obtained with the QTF. The image size is 3 $\mu$m $\times$ 3 $\mu$m.
(a)--(b) The photo-induced force measured in the shear force mode. 
(c)--(d) The photo-induced force measured when using the tapping mode. 
The yellow arrow depicts the oscillating direction of the tip while the white arrow is the electric field polarization direction (assumed to be along the $x$-axis).
Red arrows in each panel show the position where cross sections are taken, displayed below each panel. The open red dots in the cross sections show the normalized QTF amplitude ($A_{\rm{s}}\propto\partial_{s}F_{s}$), which are compared to the simulated values $\mid\partial_{i}^{2}\;\bold{p}\cdot\bold{E}\mid$ (black solid line) retrieved from Figures \ref{f6}.
}
\end{figure}

In general, measuring laterally directed photo-induced forces is challenging. The field gradient of a tightly focused beam in the lateral dimension is rather shallow on the nanometer scale, resulting in relatively weak photo-induced forces. Here we demonstrate that the sensitivity of the QTF is sufficient to register laterally directed photo-induced forces. In Figure \ref{f7}(a) and (b), we show the PiFM signal as measured through the fundamental mode when the tip is mounted as in Figure \ref{f5}(c) and thus sensitive to the shear force. The white arrows in Figure \ref{f7} indicate the direction of the incident field polarization, whereas the yellow arrows show the tip oscillating direction.  In panel \ref{f7}(a), the field polarization is aligned with the tip oscillation direction, whereby the QTF is driven by force gradient in the $x$-direction, $\partial_{x}\langle{\bf{F}}\rangle_x$.  The PiFM image shows a characteristic triple-lobe, indicative of the pattern simulated in Figure \ref{f6}(g). The measured image thus provides evidence that the shear force mode samples the gradient of the photo-induced force in the lateral ($x$) direction. 

In Figure \ref{f7}(b), the field polarization is rotated 90$^o$, whereas the tip oscillation coordinate has remained unchanged. We observe that the PiFM pattern remains largely unchanged, with a triple-lobe pattern directed predominantly along the tip oscillation axis. If we define the incident field as $x$-polarized, the tip motion then samples the gradient of the photo-induced force along the $y$-direction of the focal field, i.e. $\partial_{y}\langle{\bf{F}}\rangle_y$. In this case, the triple-lobe pattern is the expected signature. The similarity of panel \ref{f7}(a) and \ref{f7}(b) stems from the fact that the spatial distribution of the focal field in the lateral dimension is dominated by the $x$-polarized component, which is nearly rotationally symmetric. Consequently, the spatial distribution of the lateral photo-induced force and its corresponding gradient are mostly invariant upon rotating the field polarization. The red dotted line indicates the line through which a cross section is taken, shown below each panel. The open red dots in the cross-sectional graphs indicate the normalized QTF amplitude values, which are compared to computed values of $\partial_{x}\langle\bold{F}\rangle_{x}$ in \ref{f7}(a) and $\partial_{y}\langle\bold{F}\rangle_{y}$ in \ref{f7}(b), as indicated by the black solid line obtained from Figure \ref{f6}. The main features are reproduced, confirming that the shear force mode of the QTF is sensitive to lateral forces. Note that the shear force mode represents a true non-contact probing technique. At no time during the measurement is the tip in contact with the glass surface, emphasizing that the measured PiFM signal results from the direct interaction of the tip's polarizability and the local electromagnetic field, and is thus independent of tentative photothermal modulation of the glass surface.

A different observation is made when the tuning fork is mounted as in Figure \ref{f5}(b), which represents the tapping mode, making the measurement sensitive to the gradient of the force in the $z$-direction. The tapping mode benefits from the tip anisotropy, which has its largest polarizability along $z$, and is preferentially driven by the $E_{0z}$ component of the incident field. We thus expect a PiFM pattern that resembles the image shown in Figure \ref{f6}(i).  Indeed, as depicted in Figure \ref{f7}(c), the PiFM image reveals a double-lobed pattern with a clear nodal plane at the center of the focal spot. This experiment confirms that the tapping mode senses the $z$-polarized component of the field. Unlike the $E_{0x}$ component, the $z$-polarized field component is not invariant upon a rotation in the lateral plane. This is shown in Figure \ref{f7}(d), where the input polarization is rotated by 90$^o$. The double-lobed pattern remains, but is rotated by 90$^o$, as expected for a $E_{0z}$ dominated response. The latter confirms that the tapping force tracks the $z$-directed photo-induced force as given $\partial_{z}\langle\bold{F}\rangle_{z}$. These measurements thus reveal that the shear force mode and the tapping mode of the QTF probe different directional forces.  

\section{\label{sec:level1}IV. Conclusion}
In this work, we have analyzed the mechanical modes of the QTF through FEM simulations and experiments. By comparing the simulations and experiments, we were able to assign the first seven asymmetric eigenmodes of the QTF. We also presented a modified version of single beam theory, which provides a helpful guide for assigning the eigenmodes of the QTF once the dimensions of the tuning fork and the frequency of its first fundamental in-plane bending mode are known. Unlike the cantilever, the QTF provides a series of distinct eigenmodes which can be used to interrogate different aspects of the tip-sample interaction in AFM measurements. The versatility of the QTF for multi-frequency AFM was exemplified by using the tuning fork for PiFM measurements, where one mode is used for feedback while a second mode is used for monitoring the force induced by a tightly focused laser beam. The versatility of the QTF made it possible to probe the laterally and axially directed photo-induced forces independently, a capability that cannot be easily accomplished with regular cantilever beams. In addition, the shear force mode enabled by the QFT constitutes a genuine non-contact probe of laterally directed forces. We expect that the analysis and demonstration presented here will open the door to more multi-frequency applications of the QTF.

% \\\\\\\\\\\\\\\\\\\\\\\\\\\\\\\\\\\

\begin{acknowledgments}
This work was supported by the Department of Energy through grant DE-SC0013172 and by the National Science Foundation through grant CHE-1414466.
\end{acknowledgments}

%\nocite{*}

\bibliography{cite}% Produces the bibliography via BibTeX.

\end{document}